\documentclass{elsart4-1}

\usepackage{graphicx}
\usepackage{color}

\usepackage{amssymb}
\usepackage[english,francais]{babel}

\newtheorem{e-proposition}[theorem]{Proposition}

\newtheorem{e-definition}[theorem]{Definition\rm}

\renewcommand{\theequation}{\arabic{equation}}
\def\og{\leavevmode\raise.3ex\hbox{$\scriptscriptstyle\langle\!\langle$~}}
\def\fg{\leavevmode\raise.3ex\hbox{~$\!\scriptscriptstyle\,\rangle\!\rangle$}}

\begin{document}

\centerline{Physics}
\begin{frontmatter}

\selectlanguage{english}
\title{High-performance electronic cooling with superconducting tunnel junctions}


\selectlanguage{english}
\author[authorlabel1]{H. Courtois},
\ead{herve.courtois@neel.cnrs.fr}
\author[authorlabel2,authorlabel3]{H. Q. Nguyen},
\author[authorlabel1]{C. B. Winkelmann},
\author[authorlabel3]{J. P. Pekola}

\address[authorlabel1]{Universit\'e Grenoble Alpes, Institut N\'eel, F-38000 Grenoble, France and CNRS, Institut N\'eel, F-38000 Grenoble, France}
\address[authorlabel2]{Nano and Energy Center, Hanoi University of Science, VNU, Hanoi, Vietnam}
\address[authorlabel3]{Low Temperature Laboratory, Department of Applied Physics, Aalto University School of Science, FI-00076 Aalto, Finland}

\medskip
\begin{center}
{\small Received *****; accepted after revision +++++}
\end{center}

\begin{abstract}
When biased at a voltage just below a superconductor's energy gap, a tunnel junction between this superconductor and a normal metal cools the latter. While the study of such devices has long been focussed to structures of submicron size and consequently cooling power in the picoWatt range, we have led a thorough study of devices with a large cooling power up to the nanoWatt range. Here we describe how their performance can be optimized by using a quasi-particle drain and tuning the cooling junctions' tunnel barrier.
{\it To cite this article: H. Courtois, H. Q. Nguyen, C. B. Winkelmann, J. P. Pekola, C. R. Physique x (2016).}

\vskip 0.5\baselineskip

\selectlanguage{francais}
\noindent{\bf R\'esum\'e}
\vskip 0.5\baselineskip
\noindent
{\bf Refroidissement \'electronique \`a haute performance par jonction tunnel supraconductrice.}
Polaris\'ee \`a une tension juste inf\'erieure \`a la bande interdite du supraconducteur, une jonction tunnel entre ce supraconducteur et un m\'etal normal peut refroidir ce dernier. Alors que les \'etudes de ces dispositifs se sont longtemps concentr\'ees sur des structures de taille submicronique, en cons\'equence avec des puissances de refroidissement de l'ordre du picoWatt, nous avons men\'e une \'etude compl\`ete de jonctions NIS avec une forte puissance de refroidissement, de l'ordre du nanoWatt. Dans cette revue, nous d\'ecrivons comment leurs performances peuvent \^etre optimis\'ees par l'ajout d'un drain pour les quasi-particles et l'ajustement de la barri\`ere tunnel des jonctions r\'efrig\'erantes.
{\it Pour citer cet article~: H. Courtois, H. Q. Nguyen, C. B. Winkelmann, J. P. Pekola, C. R. Physique x (2016).}

 \vskip 0.5\baselineskip
\noindent{\small{\it Mots-cl\'es~:} Jonctions tunnel~; Refroidissement \'electronique~; Thermo\'electricit\'e}}
\end{abstract}
\end{frontmatter}

\selectlanguage{english}
\section{Introduction}

Due to Joule heating, an electronic bath in a circuit has usually a temperature higher than the bath temperature. The ability of electronic cooling thus opens unusual perspectives, both practical or fundamental.

In general terms, cooling an ensemble of particles can be achieved by replacing high-energy particles by low-energy ones. This selective evaporation scheme requires the implementation of an energy-selective filter. The electronic density of states of a superconductor offers such a filter as it is zero within an energy gap centered at the Fermi level. Electron tunneling through a NIS junction between a normal metal to be cooled and a superconductor is strongly energy-selective: only electrons with an energy (with respect to the Fermi level) higher than the energy gap can escape from the metal and only electrons with an energy below the energy gap can be injected into the same metal. In this way, the electronic temperature of the electronic population as a whole is reduced compared to the environment.

Assuming that the electronic populations in both the normal metal and the superconductor can be described by Fermi distributions $f_N$ and $f_S$ at respective temperatures $T_N$ and $T_S$, the cooling power of a NIS junction biased at a voltage $V$ writes \cite{GiazottoRMP06,MuhonenRPP,JLTP-Courtois}:
\begin{equation}
	\dot{Q}_{NIS} =\frac{1}{e^2R_{T}}\int^{\infty}_{-\infty}(E-eV)N_{S}(E)[f_{N}(E-eV)-f_{S}(E)]dE.
	\label{coolingpower}
\end{equation}
Here, $R_T$ is the tunnel resistance, $N_S$ is the superconductor's density of states, $k_B$ is the Boltzmann constant and $e$ is the electron charge. At the optimum cooling bias $eV\simeq\Delta-0.66k_BT_N$ and at low temperature $T_N\ll T_c$, where $T_c$ is the critical temperature, it is: 
\begin{equation}
	\dot{Q}_{NIS}\simeq 0.59 \frac{\Delta^2}{e^2R_T}\left(\frac{k_BT_N}{\Delta}\right)^{3/2},
\end{equation}
where $\Delta$ is the superconductor's energy gap. The efficiency of the cooler is:
\begin{equation}
	\eta=\frac{\dot{Q}_{NIS}}{IV}\simeq0.7\frac{T_N}{T_c},
\end{equation}
where $I$ is the (charge) current. It amounts to about $20\%$ near $T_N$ = 350 mK for aluminium with a $T_c \simeq$ 1.3 K. Aluminum is the standard choice of a superconductor thanks to the high-quality of its oxide which ensures a tunnel barrier without pinholes.

The heat current in a NIS junction \cite{NahumAPL94} is an even function of the voltage bias, which makes that a SINIS junction biased at a double bias (close to 2$\Delta$) operates just like a simple NIS junction but with a double power and most importantly a very good thermal isolation \cite{LeivoAPL96}. This makes that SINIS junctions are always preferred to plain NIS junctions.

The smaller the tunnel resistance of a junction, the larger its cooling power, as long as only single-particule tunneling is considered. The contribution of Andreev reflection to the transport can be enhanced by the confinement by disorder and lead to a quite detrimental heat current even though the charge current remains small \cite{BardasPRB95,VasenkoPRB10,RajauriaPRL08}. In terms of cooling, there is thus an optimum for the barrier transparency.

In general, the most significant opposing heat current to $\dot{Q}_{NIS}$ comes from the electron-phonon interaction in the normal metal. The most accepted form for a metal writes
\begin{equation}
	\dot{Q}_{e-ph}=\Sigma\mathcal V(T_N^5-T_{ph}^5),
	\label{phonons}
\end{equation}
where $\Sigma=2\times10^9$ WK$^{-5}$m$^{-3}$ for Cu, $\mathcal{V}$ is the volume of the normal island, $T_{ph}$ is the phonon temperature. If the phonons are weakly coupled to the external world, they can be cooled as a consequence of electron cooling. This effect was first identified through the analysis of electronic coolers' performance \cite{RajauriaPRL07,KoppinenPRL09} and then directly identified through the measurement of the phonon temperature \cite{PascalPRB13}. As will be discussed below, phonon cooling can significantly improve the performance of electronic coolers.

Still, the main limitation to electronic cooling is widely recognized as being the imperfect evacuation of quasi-particles created by the tunneling events from the vicinity of the tunnel junctions in the superconducting electrodes \cite{PekolaAPL00}. The decay of the quasiparticle density involves quasiparticle recombination retarded by phonon retrapping \cite{RothwarfPRL67,RajauriaPRB09}. In this process, two quasiparticles initially recombine to form a Cooper pair, resulting in the creation of a phonon of energy $2\Delta$. This phonon can be subsequently re-absorbed by a Cooper pair, resulting in two new quasiparticles. The $2\Delta$ energy leaves the superconductor when either the phonons or the quasiparticles escape to the bath. The basic strategy to address this issue relies on the presence of quasi-particles traps made of pieces of normal metal coupled to the superconducting electrodes \cite{VoutilainenPRB00,CourtPRB08,WangNatComm14}. Still, the poor coupling of the traps due to the presence of the same tunnel barrier as in the cooling junctions is a severe limitation. As an interesting alternative, vortices could be created in the superconducting electrodes by applying a magnetic field, and their position was controlled by the geometry of the electrodes \cite{PeltonenPRB10}. Vortices act as a local trap for quasi-particles but this approach imposes a magnetic field, which is not compatible with the use of large tunnel junctions.

Eventually, the practical implementation of electronic cooling \cite{MillerAPL08,LowellAPL13} actually calls for large cooling powers, well above the usual picoWatt range of electronic coolers fabricated with angle evaporation on a suspended resist mask. As the charge current is larger in this case, a large-area junction brings more stringent conditions for the quasi-particle evaporation \cite{ONeilPRB12}. The geometry, the quasi-particle evacuation and the tunnel barrier transparency need to be specifically optimized.

In this paper, we review our recent work on electronic coolers based on large-area NIS junctions. We describe a strategy which enabled us to reach unprecedented performances by solving to a great extent the long-standing question of evacuation of the quasi-particles generated by the cooler's operation.

\section{Implementation of superconducting coolers from a multilayer}

\begin{figure}[t]
\begin{center}
\includegraphics[width=0.8\columnwidth,keepaspectratio]{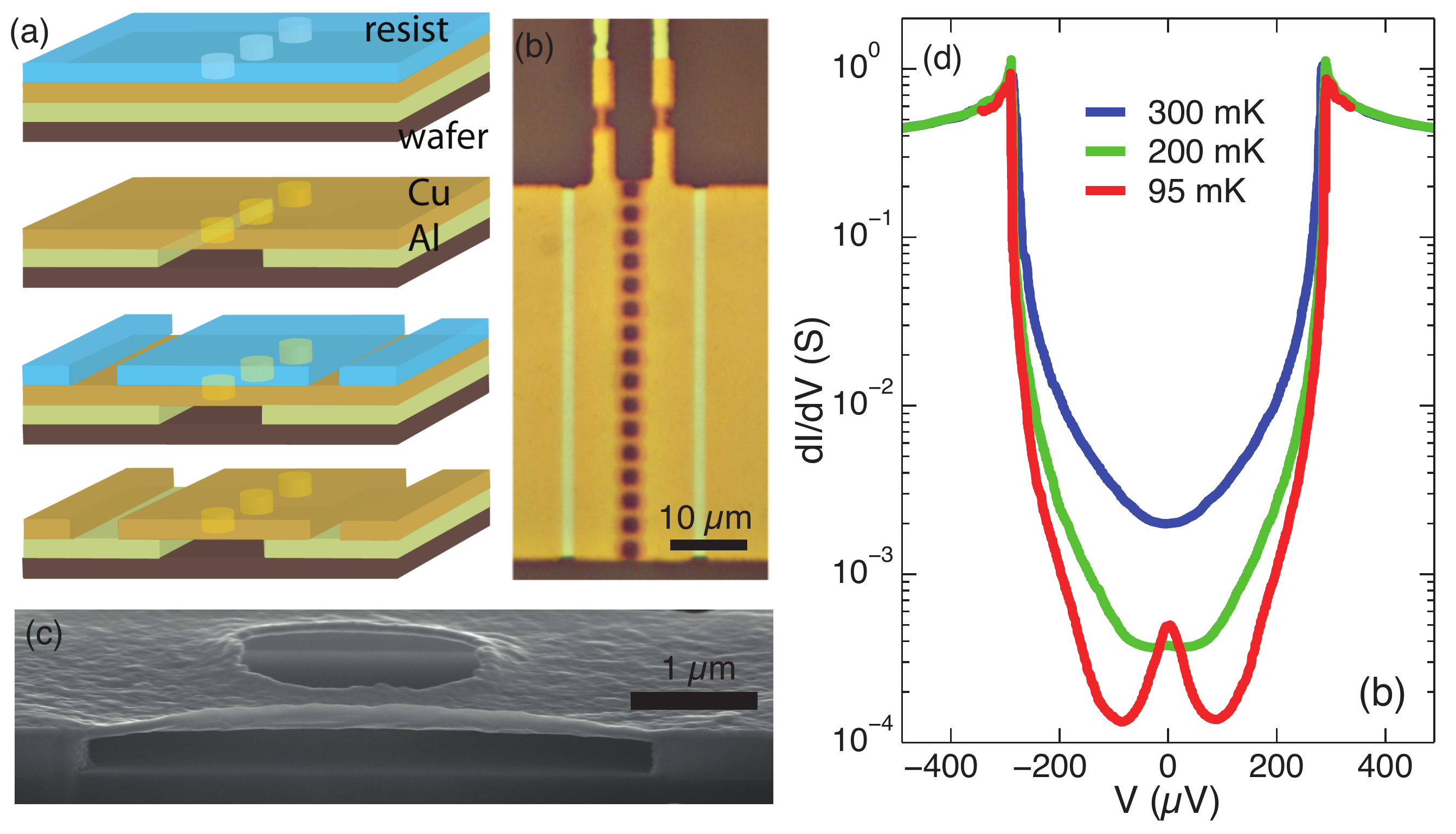}
\caption{(Color online.) (a) From top: fabrication starts with an Al/AlOx/Cu multilayer, on which a photoresist (blue color) is patterned with contact pads and holes. Then, Cu (orange) and Al (green) layers are successively etched, leaving a suspended membrane of Cu along the line of adjacent holes. A second lithography and etch define the Cu central island. (b) Optical microscope image showing regions by decreasing brightness: bare Al, Cu on Al, suspended Cu and substrate. On the top, two thermometer junctions are added. (c) Scanning electron micrograph of a sample cut using Focused Ion Beam, showing the Cu layer suspended over the holes region. (d) Differential conductance of sample A at different bath temperatures.}
\label{CRP}
\end{center}
\end{figure}

We have developed a new approach for the fabrication of large-area tunnel junctions \cite{NguyenAPL}. A wafer-scale multilayer is selectively etched either isotropically or anisotropically by using masks patterned by UV lithography. Figure \ref{CRP}a gives an overview of the process in the simple case of a Al/AlOx/Cu multilayer. A first UV lithography of a resist layer (blue color) defines the overall geometry of the device with its contacts, as well as a series of holes (or dots), forming a kind of dotted line. The Cu layer (orange) is first etched with an Ar ion beam or using chemicals. The Al layer (green) is afterwards isotropically etched with a weak base. The dotted line was designed so that, thanks to the lateral over-etch, the etched regions around every dot overlap, thus forming two separate Al electrodes. The Cu central island is afterwards defined through a second UV lithography and etch. Figure \ref{CRP}b optical image shows a complete device while Figure \ref{CRP}c shows a side-view of a sample cut with a focussed ion beam. This process provides SINIS junctions with absolutely no limitation in area and a structural quality determined by the initial multilayer which can be epitaxial.

\begin{table}[t]
	\begin{center}
\begin{tabular}{@{}ccccccc}\hline
Samples & Drain & Cooler & Direct & $2R_T$ & $2\Delta$ & Figure\\
& barrier & barrier & trap & $(\Omega)$ & $(\mu$eV)&\\
\hline
A & N/A & N/A & no &  2.8 & 360 & 1 \\ 
B1 & N/A & N/A & no &  0.83 & 390 & 2 \\ 
B2 & N/A & N/A & yes &  0.61 & 390 & 2 \\ 
C & 0 & 1.3,180 & no &  1.31 & 180 & 3 \\ 
D & 5$\times$10$^{-4}$,10 & 1.3,300 & no &  1.01 & 228 & 3 \\ 
E1 & 1.3,10 & 1.3,300 & no &  0.71 & 398 & 3 \\ 
E2 & 0.26,10 & 1.3,300 & no &  1.56 & 382 & 3 \\ 
E3 & 0.18,1 & 0.8,180 & no &  0.55 & 370 & 3 \\ 
F & N/A & 1,300 & no &  0.83 & 390 & 3 \\ 
G & 0.02*,120 & 1.3,300 & no &  1.7 & 375 & 4 \\ 
H & 0.02*,120 & 9,300 & no &  3.7 & 375 & 4 \\ 
J1 & 0.02*,120 & 13,300 & no &  4.8 & 375 & 4 \\ 
J2 & 0.02*,120 & 13,300 & yes &  4.8 & 375 & 4 \\ 
I & 0.02*,120 & 50,7200 & no &  10.5 & 375 & 4 \\ 
\hline
\end{tabular}
\caption{Devices presented throughout the paper. The drain barrier is the thin insulator between Al and AlMn, and the cooler barrier is the main barrier between Al and Cu. The related numbers refer to oxidation pressure in mbar and time in second, while the symbol $*$ denotes the use of a oxidation mixture of Ar:O$_2$ with ratio 10:1. The direct trap column notifies the presence of a direct trap or not. 2$R_T$ is the tunnel resistance of the two NIS cooling junctions measured in series, and 2$\Delta$ is the energy gap obtained from a BCS fit. "Figure" indicates the figure where data on a given sample is shown. Every sample has NIS junctions' size of 70$\times$4 $\mu$m$^2$.}
\end{center}
\end{table}

\begin{figure}[b]
\center
\includegraphics[width=0.99\columnwidth,keepaspectratio]{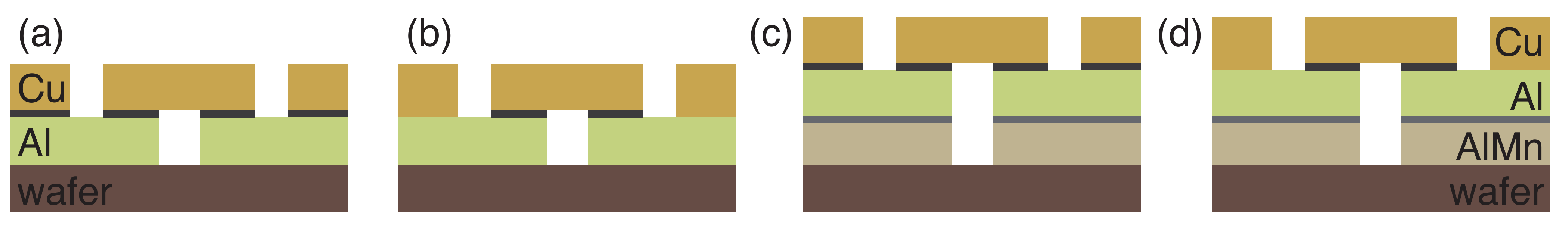}
\caption{(Color online.) Schematics of the different geometries studied in this work. (a) original geometry for samples A, B1 and F, (b) with direct traps for sample B2 (c), or with quasi-particle drains for samples C, D, E1, E2, E3, G, H, J1, and I (d), with both in sample J2.}
\label{Dessins}
\end{figure}

In the same sample, additional superconducting tunnel probes can be connected to the cooled metal. As the sub-gap (charge) current of such a NIS junction is highly sensitive to the electronic temperature, it provides a sensitive electron thermometer \cite{NahumAPL93}. The usual method is to bias the junction at a fixed and small current, chosen so that it does not contribute to cooling. The measured voltage is directly related to the electronic temperature. A high sensitivity, typically better than 0.1 mK, is easily achieved.

Figure \ref{CRP}d shows the differential conductance of a typical device at various bath temperatures on a logarithmic scale which highlights the details of the sub-gap conductance. The typical behavior of a SINIS junction is obtained, with in addition a differential conductance peak at zero bias arising from Andreev reflection. At this point, one can distinguish the presence of electronic cooling through the curvature of the differential conductance plot in the sub-gap regime. An isothermal behavior would indeed exhibit a linear dependence on a semi-log scale. Still, the electronic cooling remains quite modest, for instance by 60 mK starting from a bath temperature of 300 mK \cite{NguyenAPL}.

\section{Direct trap}

\begin{figure}[t]
\center
\includegraphics[width=0.72\columnwidth,keepaspectratio]{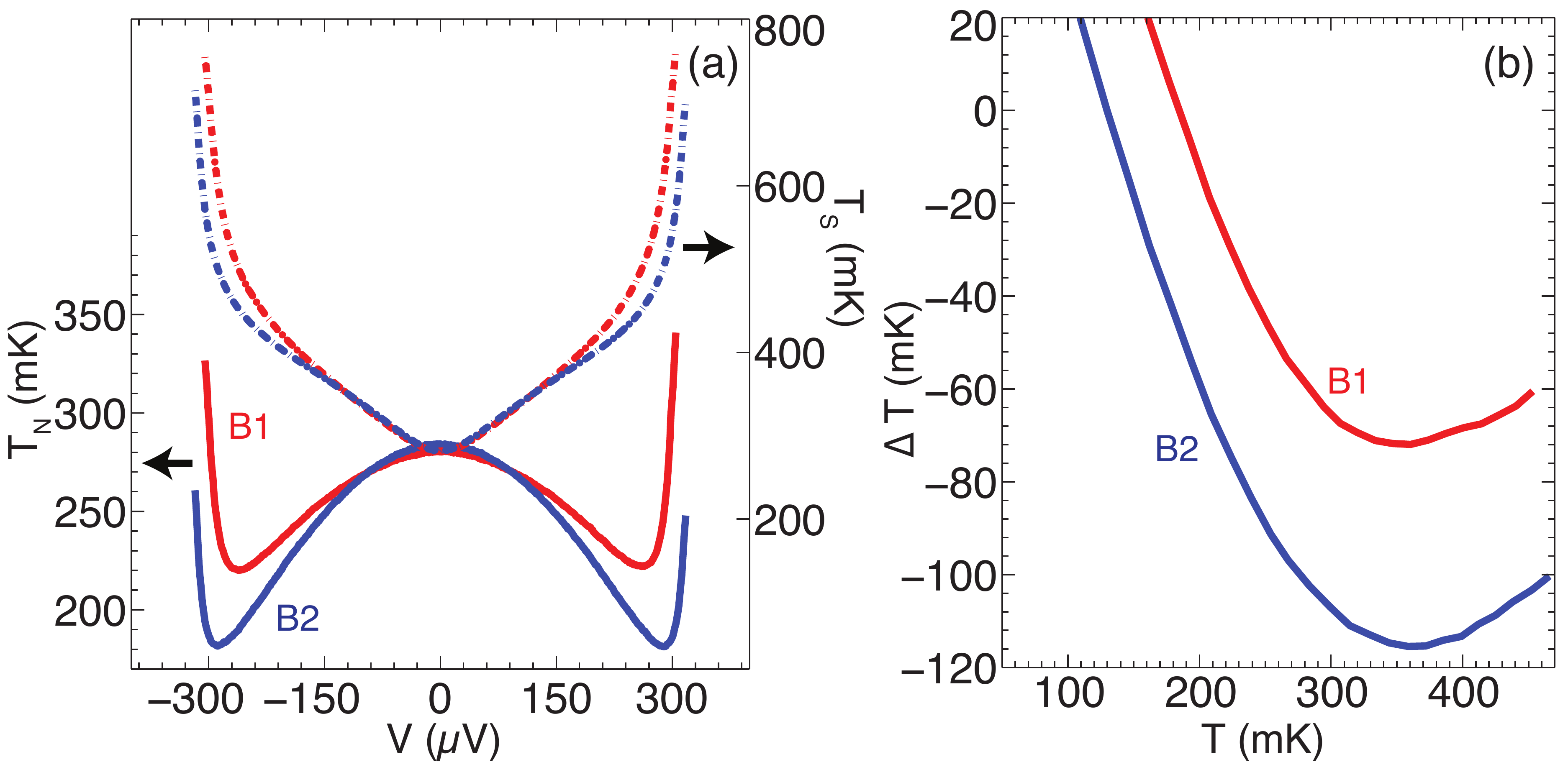}
\caption{(Color online.) (a) Calculated electron temperature (dashed lines, right axis) of the superconductor and measured electron temperature of the normal metal (solid lines, left axis) for the same sample with tunnel-coupled traps (B1, red lines) or with direct traps (B2, blue line) as a function of the bias voltage. (b) Temperature drop at the optimum bias as a function of the bath temperature.}
\label{Direct}
\end{figure}

Owing to the large junction area of our devices, the estimated cooling power of devices such as described above lies in the nanoWatt range. Still, the related large bias current makes the question of quasiparticle evacuation much more stringent than in conventional devices, which can limit the performance. In the original design, see Fig. \ref{Dessins}a, quasiparticle traps are naturally present since a Cu layer covers the Al electrodes. Nevertheless, this layer is coupled to the electrodes through a tunnel barrier. 

As a first step, we have replaced these tunnel-coupled traps by direct traps, \mbox{i.e.} metal directly coupled to the superconducting electrodes, without oxide in-between, see Fig. \ref{Dessins}b. This is done by a separated lithography, where the Cu lateral traps are etched. The AlOx barrier is afterwards removed by Ar plasma in vacuum, which is followed by Cu deposition and lift-off. This direct contact between Al and Cu should both evacuate efficiently quasiparticles from the electrodes and couple them more strongly to the bath.

Figure \ref{Direct} compares the behavior of the same sample in its original state with traps coupled to the leads through a tunnel barrier (B1), and after subsequent modification with direct traps (B2). A significant improvement of the cooling performance is observed, with the lowest temperature achieved dropping from 275 down to 229 mK. It is related to a significant drop of the superconducting leads' electronic temperature, which nevertheless remains quite high. The obtained performance thus remains far from the theoretical expectation in the hypothesis of efficient evacuation of quasiparticles. We conclude that direct traps alone cannot solve the issue of quasiparticles evacuation for these high power coolers.

\section{Quasi-particle drain}

The efficiency of lateral traps is limited by the distance between the injection regions and the traps, up to a few microns in our case. In order to address this, we have modified the device geometry by adding another normal metal layer below the superconducting Al electrodes \cite{NguyenNJP,NguyenPRAppl}, see Fig. \ref{Dessins}c. This additional layer acts as a quasiparticle drain. As for the material, an AlMn alloy \cite{ClarkAPL04} was chosen as it retains the Al oxide quality while being non-superconducting. More importantly, it also carries the same chemical properties as Al during the chemical etch, so that the two layers etched simultaneously yield an identical final geometry. The quasiparticle drain stays so close to the NIS junction that an oxide layer is required to stop proximity effect that can soften the superconducting gap. This tunnel interface between the AlMn drain and the Al electrodes can be tuned independently of the tunnel barrier of the cooling junction, thus bringing additional flexibility for the device optimization. 

\begin{figure}[t]
\begin{center}
\includegraphics[width=0.86\columnwidth,keepaspectratio]{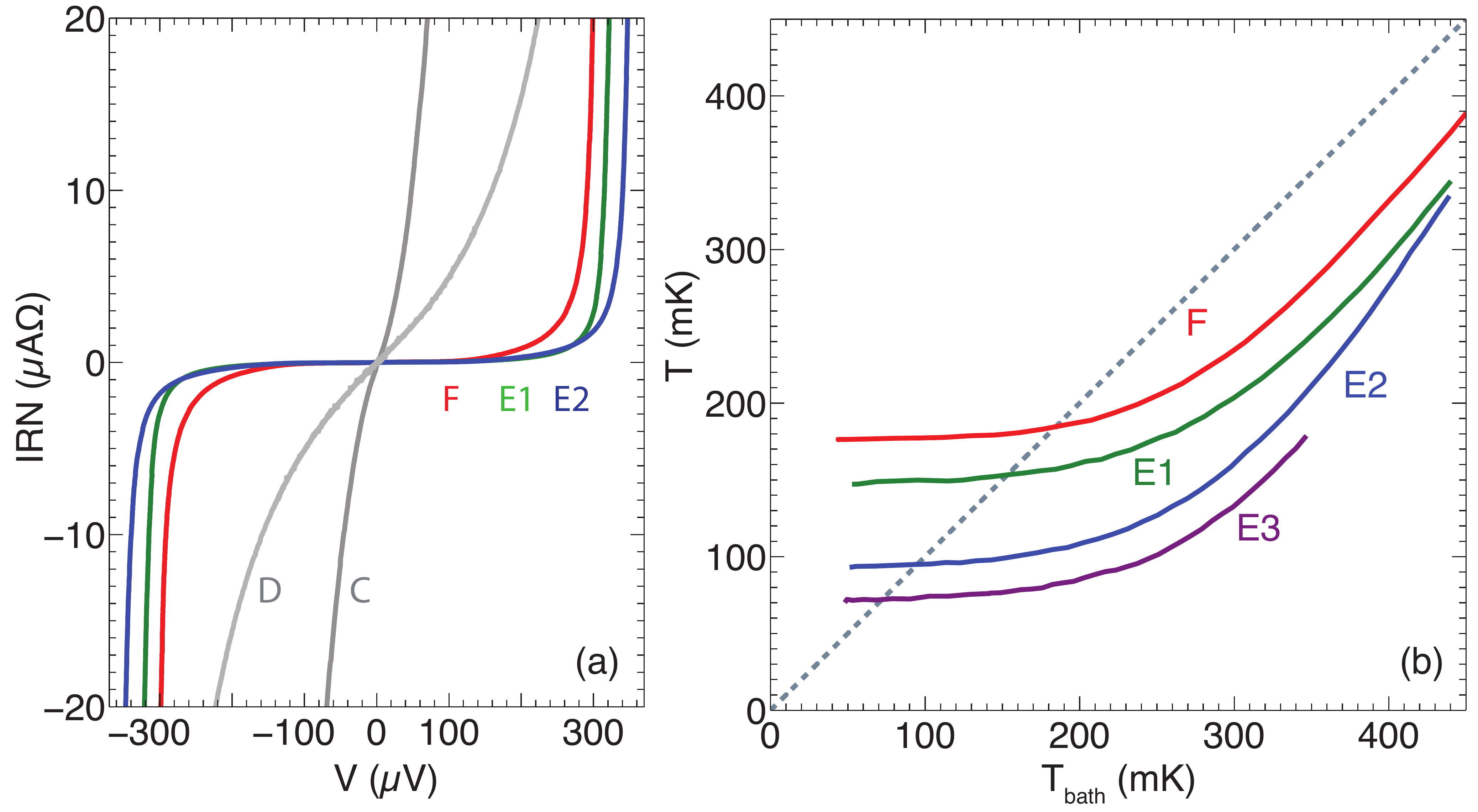}
\caption{(Color online.) (a) Current-voltage characteristic at a 50 mK bath temperature of samples C-F with different barrier thicknesses between the quasi-particles' drain and the superconducting leads, see Table 1. (b) Temperature of the normal metal island $T_N$ at the optimum bias as a function of bath temperature $T_{bath}$. Samples E1, E2, and E3 differ only in their tunnel resistances between the drain and the superconducting electrodes (see Table 1). The gray dotted line is the 1-1 line at the boundary between cooling and heating.}
\label{Drain}
\end{center}
\end{figure}

Figure \ref{Drain} a shows the current voltage characteristic at a bath temperature of 50 mK for a series of samples with different drain barrier transparency, but (almost) identical barriers for the cooling junctions, see Table 1. The two innermost curves stand for sample C and D, which have no barrier at the drain interface with the leads or a very thin barrier, respectively. Superconductivity in the Al electrodes is then affected by a strong inverse proximity effect, which results in poor electronic cooling. Samples E1, E2, E3 are fabricated with a stronger barrier for the drain. They show a sharp characteristic, with sample E3 (not shown) behaving very similarly to sample E2. Figure \ref{Drain} b displays the electronic temperature achieved at the optimum point as a function of the bath temperature. The presence of a quasi-particle drain thus improves significantly the cooling performance. Sample E3 with the minimum oxidation on the drain tunnel junction shows the best cooling among the sample set.

\begin{figure}[t]
\begin{center}
\includegraphics[width=\columnwidth,keepaspectratio]{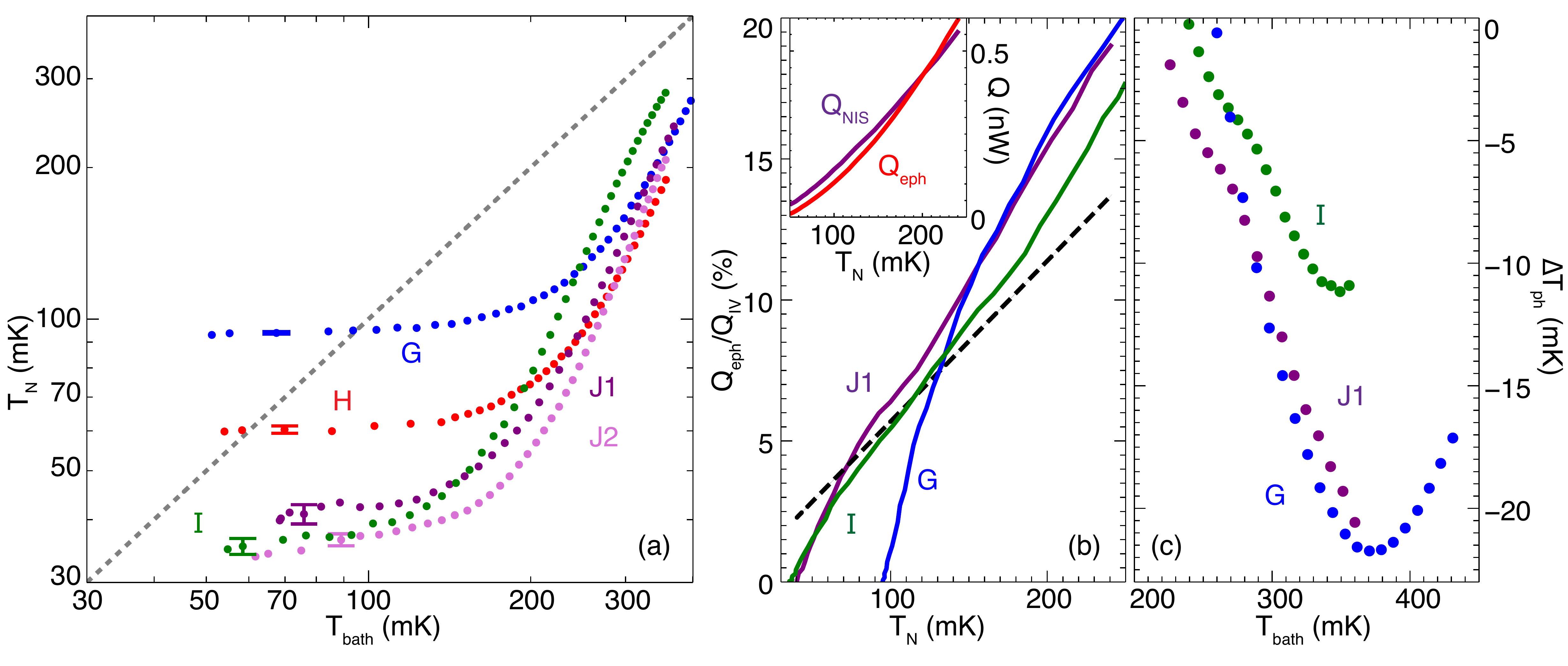}
\caption{(Color online.) (a) Temperature of the normal metal island $T_N$ at the optimum bias as a function of bath temperature $T_{bath}$. Samples G, H, I, and J1 differ only in their tunnel resistances $R_T$ (Table 1). J2 is an improved version of J1, see text. The gray dotted line is the 1-1 line at the boundary between cooling and heating. (b) Apparent efficiency with the assumption of metal phonons thermalized at the bath temperature for samples G, I and J1 compared to the theory prediction Eq. (2) (black dashed line). The inset shows the calculated $\dot{Q}_{NIS}$ when assuming $T_S$ = $T_{bath}$ and $\dot{Q}_{e-ph}$ when assuming $T_{ph}$ = $T_{bath}$ for sample J1. (c) Extracted phonon temperature of the normal island $\Delta T_{ph} = T_{ph} - T_{bath}$ assuming the theoretical efficiency and no over-heating of the leads.}
\label{Drain2}
\end{center}
\end{figure}

Let us now consider the effect of the resistance of the cooling junctions on the cooling performance. A smaller tunnel resistance leads to a large cooling power, which is beneficial, but also to a stronger quasiparticle injection, which is detrimental. Figure \ref{Drain2} shows the electronic temperature at the optimum point for a series of samples differing in this respect only. Over our sample set, sample J2 shows the best compromise between performance at very low temperature and operation at higher temperature. It reaches a record electronic temperature of 30 mK when the bath temperature is 150 mK. This achievement confirms the relevance of the quasi-particle drain geometry for the evacuation of quasi-particles generated by the cooler's operation.

Made from sample J1, the optimized sample J2 has direct traps (Fig. \ref{Dessins}d) and is measured with the highest shielding possible. The moderate improvement from J1 to J2 again confirms that direct traps do not provide a dominant relaxation mechanism. Compared to J1-2, samples G and H have a thinner drain barrier and do not perform as well at low temperature. Sample I has a thicker barrier and performs equally well as J1-2 at low temperature but not as well in the intermediate temperature regime.

When comparing the calculated cooling power $\dot{Q}_{NIS}$ from Eq. (\ref{coolingpower}) to the electron-phonon coupling power $\dot{Q}_{eph}$ from Eq. (\ref{phonons}), one concludes that  the phonons are not well thermalized at the bath temperature, but cool to a lower intermediate temperature, \mbox{i.e.} $T_{ph}<T_{bath}$. Our present device geometry is actually quite relevant for phonon cooling, as the cooled normal metal is isolated from the substrate. If one assumes the existence of independent phonon populations, the phonons of the cooled metal are coupled to the bath through the superconducting electrodes, which introduces at least two interfaces between different materials. The related Kapitza resistance thus significantly decouples the metal phonons from the bath, which enhances electronic cooling. The phonon cooling extracted from the data analysis amounts up to 20 mK \cite{NguyenPRAppl}. It is maximum at relatively high temperatures of about 350 mK, which is expected since the Kapitza resistance is proportional to $T^4$ while the electron-phonon coupling varies as $T^5$.

\section{Conclusion}

We have shown how electronic cooling can be optimized in specially-designed Normal metal - Insulator - Superconductor junctions with a large area. The two key ingredients that have been worked on are: (i) the tunnel barrier transparency for the cooling junctions, (ii) the coupling to a quasiparticle drain, again through a (separate) tunnel junction. This being done, we have demonstrated a temperature reduction of a factor 5, from 150 mK down to 30 mK, and a cooling power of the order of one nanoWatt. Further improvement could be achieved by using active traps, where the traps themselves are cooled electronically \cite{HungPreprint}, or, similarly, cascade coolers where the superconducting electrodes of a SIN device are directly cooled using a SIS' junction, where S' is a superconductor with a larger energy gap \cite{CamarasaAPL14}. Moreover, spin-filtering barriers may help in eliminating Andreev processes \cite{APL-Vasenko}.

\section*{Acknowledgements}
We have benefited from discussions with and contributions from M. Meschke, J. T. Peltonen, F. W. J. Hekking, F. Giazotto and A. Vasenko. We acknowledge the support of the Nanoscience Foundation - Grenoble.


\begin{thebibliography}{00}
\bibitem{GiazottoRMP06} F. Giazotto, T. T. Heikkil\"a, A. Luukanen, A. M. Savin and J. P. Pekola, {\bf Opportunities for mesoscopics in thermometry and refrigeration: physics and applications}, Rev. Mod. Phys. \textbf{78}, 217 (2006).
\bibitem{MuhonenRPP} J. T. Muhonen, M. Meschke, and J. P. Pekola, {\bf Micrometre-scale refrigerators}, Rep. Prog. Phys. \textbf{75}, 046501 (2012).
\bibitem{JLTP-Courtois} H. Courtois, F. W. J. Hekking, H. Q. Nguyen and C. B. Winkelmann, {\bf Electronic Coolers Based on Superconducting Tunnel Junctions: Fundamentals and Applications}, Journal of Low Temperature Physics {\bf 175}, 799 (2014).
\bibitem{NahumAPL94} M. Nahum, T. M. Eiles and J. M. Martinis, {\bf Electronic microrefrigerator based on a normal-insulator-superconductor tunnel junction}, Appl. Phys. Lett. \textbf{65}, 3123 (1994).
\bibitem{LeivoAPL96} M. M. Leivo, J. P. Pekola, and D. V. Averin, {\bf Efficient Peltier refrigeration by a pair of normal metal/insulator/ superconductor junctions}, Appl. Phys. Lett. \textbf{68}, 1996 (1996).
\bibitem{BardasPRB95} A.~Bardas and D.~Averin, {\bf Peltier effect in normal-metal-superconductor microcontacts}, Phys. Rev. B {\bf 52}, 12873 (1995).
\bibitem{VasenkoPRB10} A. S. Vasenko, E. V. Bezuglyi, H. Courtois, F. W. J. Hekking, {\bf Electron cooling by diffusive normal metal-superconductor tunnel junctions}, Phys. Rev. B {\bf 81}, 094513 (2010).
\bibitem{RajauriaPRL08} S. Rajauria, P. Gandit, T. Fournier, F. W. J. Hekking, B. Pannetier, and H. Courtois, {\bf Andreev Current-Induced Dissipation in a Hybrid Superconducting Tunnel Junction}, Phys. Rev. Lett. \textbf{100}, 207002 (2008).
\bibitem{RajauriaPRL07} S. Rajauria, P. S. Luo, F. W. J. Hekking, H. Courtois, and B. Pannetier, {\bf Electron and phonon cooling in a superconductor - normal metal - superconductor tunnel Junction}, Phys. Rev. Lett. \textbf{99}, 047004 (2007).
\bibitem{KoppinenPRL09} P. J. Koppinen and I. J. Maasilta, {\bf Phonon cooling of nanomechanical beams with tunnel junctions}, Phys. Rev. Lett. \textbf{102}, 165502 (2009).
\bibitem{PascalPRB13} L. M. A. Pascal, A. Fay, C. B. Winkelmann, and H. Courtois, {\bf Existence of an independent phonon bath in a quantum device}, Phys. Rev. B \textbf{88}, 100502 (2013).
\bibitem{PekolaAPL00} J. P. Pekola, D. V. Anghel, T. I. Suppula, J. K. Suoknuuti, A. J. Manninen, and M. Manninen, {\bf Trapping of quasiparticles of a nonequilibrium superconductor}, Appl. Phys. Lett. {\bf 76}, 2782 (2000).
\bibitem{RothwarfPRL67} A. Rothwarf and B. N. Taylor, {\bf Measurement of recombination lifetimes in superconductors}, Phys. Rev. Lett. {\bf 19}, 27 (1967).
\bibitem{RajauriaPRB09} S. Rajauria, H. Courtois, and B. Pannetier, {\bf Quasiparticle-diffusion-based heating in superconductor tunneling microcoolers}, Phys. Rev. B \textbf{80}, 214521 (2009).
\bibitem{VoutilainenPRB00} J. Voutilainen, T. T. Heikkil\"a, and N. B. Kopnin, {\bf Nonequilibrium phenomena in multiple normal-superconducting tunnel heterostructures}, Phys. Rev. B {\bf 72}, 054505 (2005).
\bibitem{CourtPRB08} N. A. Court, A. J. Ferguson, R. Lutchyn, and R. G. Clark, {\bf Quantitative study of quasiparticle traps using the single-Cooper-pair transistor}, Phys. Rev. B \textbf{77}, 100501 (2008).
\bibitem{WangNatComm14} C. Wang, Y. Y. Gao, I. M. Pop, U. Vool, C. Axline, T. Brecht, R. W. Heeres, L. Frunzio, M. H. Devoret, G. Catelani, L. I. Glazman and R. J. Schoelkopf, {\bf Measurement and control of quasiparticle dynamics in a superconducting qubit}, Nature Comm. {\bf 5}, 5836 (2014).
\bibitem{PeltonenPRB10} J. T. Peltonen, J. T. Muhonen, M. Meschke, N. B. Kopnin, and J. P. Pekola, {\bf Magnetic-field-induced stabilization of non-equilibrium superconductivity in a normal-metal/insulator/superconductor junction}, Phys. Rev. B \textbf{84}, 220502 (2011).
\bibitem{MillerAPL08} N. A. Miller, G. C. O'Neil, J. A. Beall, G. C. Hilton, K. D. Irwin, D. R. Schmidt, L. R. Vale, and J. N. Ullom, {\bf High resolution x-ray transition-edge sensor cooled by tunnel junction refrigerators}, Appl. Phys. Lett. \textbf{92}, 163501 (2008).
\bibitem{LowellAPL13} P. J. Lowell, G. C. OÕNeil, J. M. Underwood, and J. N. Ullom, {\bf A nanoscale refrigerator for macroscale objects}, Appl. Phys. Lett. 102, 082601 (2013).
\bibitem{ONeilPRB12} G. C. O'Neil, P. J. Lowell, J. M. Underwood, and J. N. Ullom, {\bf Measurement and modeling of a large-area normal-metal/insulator/superconductor refrigerator with improved cooling}, Phys. Rev. B \textbf{85}, 134504 (2012).
\bibitem{NguyenAPL} H. Q. Nguyen, L. M. A. Pascal, Z. H. Peng, O. Buisson, B. Gilles, C. B. Winkelmann, and H. Courtois, {\bf Etching suspended superconducting tunnel junctions from a multilayer}, Appl. Phys. Lett. \textbf{100}, 252602 (2012).
\bibitem{NahumAPL93} M. Nahum and John M. Martinis, {\bf Ultra-sensitive hot electron microbolometer}, Appl. Phys. Lett. {\bf 63}, 3075 (1993).
\bibitem{NguyenNJP} H. Q. Nguyen, T. Aref, V. J. Kauppila, M. Meschke, C. B. Winkelmann, H. Courtois and J. P. Pekola, {\bf Trapping hot quasi-particles in a high-power superconducting electronic cooler}, New J. Phys. \textbf{15}, 085013 (2013).
\bibitem{NguyenPRAppl} H. Q. Nguyen, M. Meschke, H. Courtois, and J. P. Pekola, {\bf Sub-50 mK electronic cooling with large-area superconducting tunnel junctions}, Phys. Rev. Appl. \textbf{2}, 054001 (2014).
\bibitem{ClarkAPL04} A. M. Clark, A. Williams, S. T. Ruggiero, M. L. van den Berg, and J. N. Ullom, {\bf Practical electron-tunneling refrigerator}, Appl. Phys. Lett. {\bf 84}, 625 (2004).
\bibitem{HungPreprint} H. Q. Nguyen, J. T. Peltonen, M. Meschke, J. P. Pekola, {\bf A cascade electronic refrigerator using superconducting tunnel junctions}, arXiv:1605.00830.
\bibitem{CamarasaAPL14} M. Camarasa-Gomez, A. Di Marco, F. W. J. Hekking, C. B. Winkelmann, H. Courtois, and F. Giazotto, {\bf Superconducting cascade electron refrigerator}, Appl. Phys. Lett. {\bf 104}, 192601 (2014).
\bibitem{APL-Vasenko} S. Kawabata, A. Ozaeta, A. S. Vasenko, F. W. J. Hekking and F. S. Bergeret, {\bf Efficient electron refrigeration using superconductor/spin-filter devices}, Appl. Phys. Lett. {\bf 103}, 032602 (2013).
\end{thebibliography}
\end{document}